\documentclass[conference]{IEEEtran}
\IEEEoverridecommandlockouts
\usepackage{cite}
\usepackage{amsmath,amssymb,amsfonts}
\usepackage{algorithmic}
\usepackage{graphicx}
\usepackage{textcomp}
\usepackage{xcolor}
\def\BibTeX{{\rm B\kern-.05em{\sc i\kern-.025em b}\kern-.08em
    T\kern-.1667em\lower.7ex\hbox{E}\kern-.125emX}}

\usepackage{lipsum}
\usepackage{algorithm,algorithmic}
\usepackage{multicol}

\begin{document}

\title{A Survey on Biometrics Authentication\\
%{\footnotesize \textsuperscript{*}Note: Sub-titles are not captured in Xplore and
%should not be used}
%\thanks{Identify applicable funding agency here. If none, delete this.}
}

\author{
\IEEEauthorblockN{Fangshi Zhou}
\IEEEauthorblockA{\textit{Department of Computer Science} \\
\textit{University of Dayton}\\
Dayton, OH 45469-0280\\
zhouf4@udayton.edu}

\and

\IEEEauthorblockN{Tianming Zhao}
\IEEEauthorblockA{\textit{Department of Computer Science} \\
\textit{University of Dayton}\\
Dayton, OH 45469-0280\\
tzhao1@udayton.edu}
}

\maketitle

%\begin{IEEEkeywords}
%component, formatting, style, styling, insert
%\end{IEEEkeywords}

\begin{abstract}
Nowadays, traditional authentication methods are vulnerable to face attacks that are often based on inherent security issues. Professional attackers leverage adversarial offenses on the security holes, such as unauthorized privileges after the user’s default setting, logging into the system, or seizing authorization data through eavesdropping. Biometrics has intrinsic advantages to overcome the traditional authentication methods on security, success rates, efficiency, and accessibility. Biometrics has wide prospects to implement various applications in fields. Whether in authentication security or clinical medicine, biometrics is one of the mainstream studies. In this paper, we surveyed and reviewed some related studies of biometrics, which are outstanding and significant in driving the development and popularization of biometrics. Although they still have some inherent disadvantages to restrict popularization, these obstacles could not conceal the promising future of biometrics. Multi-factors continuous biometrics authentication has become the mainstream trend of development. We reflect the findings as well as the challenges of the studies in the survey paper.
\end{abstract}

\section{Introduction}
Most cases, an ideal biometric requires several key features. First, It should be universal. It means the biometric can be applied on most users or applications. Current technologies of biometric cannot promise to apply to every user in practice. Thus, we must improve the adaptability of biometrics which can be applied to most users. The second key feature of a successful biometric is high distinguishing capability. Some kinds of biometric are employed on sensitive security applications or monitoring human health. High distinguishing certainty of biometric should be distinctly necessary and important, otherwise, the products are not only shifted out from the markets, but also incur serious consequences and loss of life and property. The third key feature of biometric is permanent. It means that biometric devices can efficiently and successfully observe and collect critical and physical features of users. These features cannot be changed and removed in most cases. However, in reality, some key physical characteristics are not really permanent and monotonous. Thus, we hope the ideal biometric adopts the features that can remain stable and changeless as long as possible in practice. The fourth key feature of ideal biometric is collectability. It means the biometric should be capable of efficiently collecting required characteristics and data from users.
Besides, regular biometric is hard to hedge high production cost. Higher accuracy of authentication often means higher production costs. Once the biometric is unfortunately cracked, the cost is high to bear. Thus, I hope to research in this direction about the biometric which could be used efficiently and economically to implement high-accuracy authentication or detection. Clearly, There are lots of newly developed technologies which have been used in biometric.

For example, no one can deny the potential and strength of machine learning nowadays. I noticed some biometric research papers of biometric have explored and improved the corporation of machine learning and biometric. Machine learning can greatly improve efficiency and reduce the cost of data collection. Because of the importance of data collection, some papers are directly dedicated to improving data acquisition techniques instead of expensive traditional sensors, such as PPG. It is a technique that verifies users via cardiac signals from common wrist-worn wearable devices.

\section{General Survey of papers}

\cite{zhao2020trueheart} proposes a low-cost continuous user authentication (CA) system, TrueHeart, that could periodically verify the identity of a target user through unique cardiac signals from a portable wrist-worn wearable device. The highlight of the proposed approach is the wearable device which provides a non-intrusive manner. It is significant for time-sensitive tasks and promises the authentication will not disturb running tasks. Besides, the paper is attractive because of its low-cost implementation. Compared to expensive dedicated sensors, such as ECG or Doppler radar sensors, the wrist-worn wearable device has an overwhelming advantage in production cost. If the technique could be mature with an extremely high success rate of authentication, it may be transplanted to clinic medicine for cardiovascular monitoring.

\begin{figure}[H]
\centering
\includegraphics[width=0.4\textwidth]{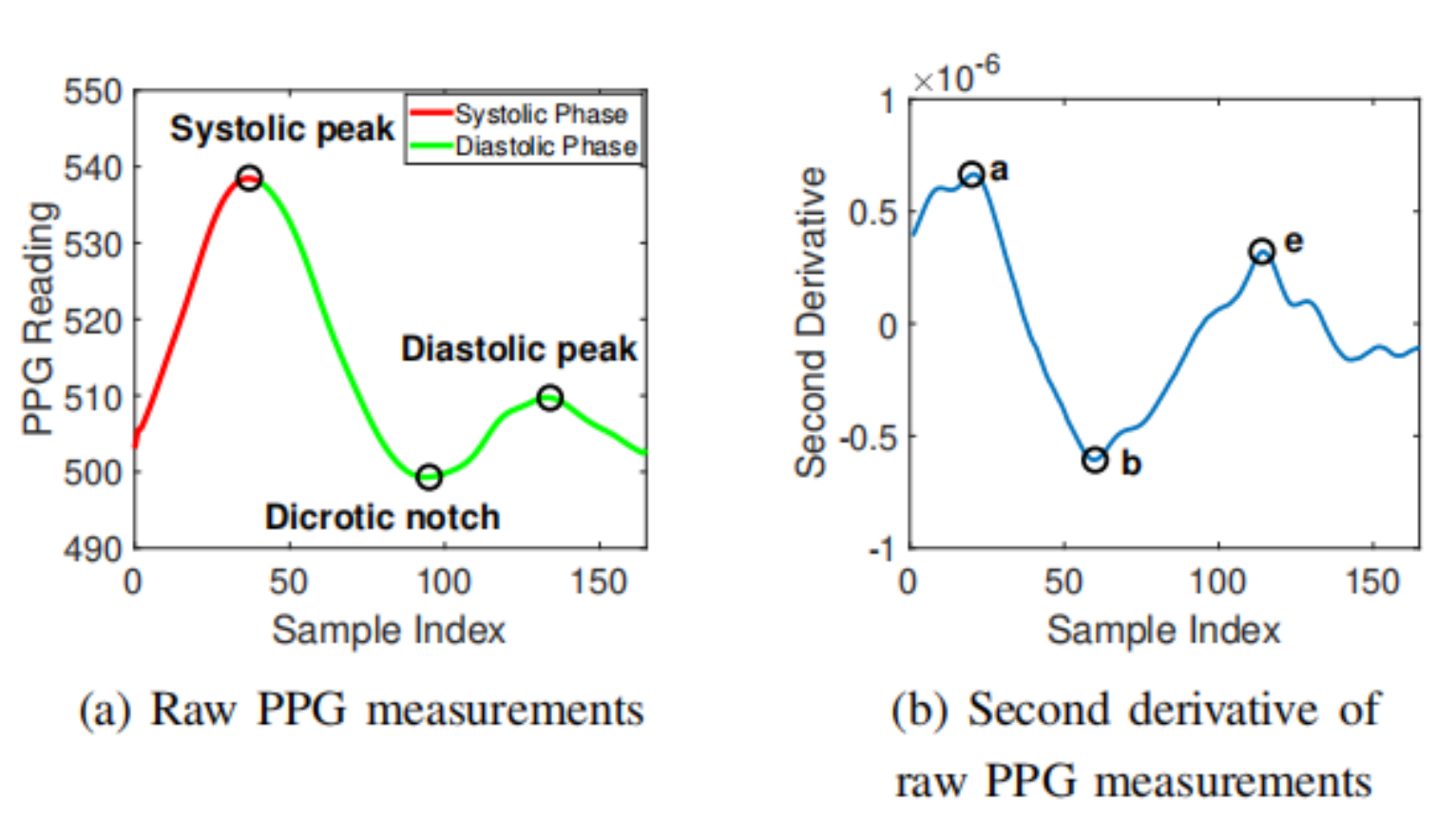}
\caption{Illustration of the critical landmarks in raw PPG measurements and its second derivative.}
\end{figure}

The paper acknowledged the proposed technique met some challenges. PPG signals are relatively coarse-grained, noisy, and more easily distorted than EGG signals. To solve these problems, the paper investigated various motion artifacts (MAs) of different types of body-movements, and introduced effective MA detection to promise to extract correct cardiac signals in practical scenarios. Besides, the paper considered several possible defensive scenes when the biometric is under attack. It shows that the biometric of authentication could be greatly improved to reach the same level of traditional sensors in some specific monitoring fields. The prerequisite of success is a mature technique of data collection, such as monitoring PPG in the paper, and considering all possible adversarial attacks and environmental interference.

\cite{ackerson2021applications} lists the application and development of Recurrent Neural Networks (RNNs) in biometric fields. As one of the most significant models in deep learning, RNNs are unique because their outputs from previous steps are fed as input to the current step of neural networks. Previously, RNNs were often applied in language generative tasks and speech recognition. Now, this paper inspired people to spread the mighty model to biometrics. Biometric authentication has been used in smart phones by reading fingerprints or scanning iris, but their application is much more than that. One of the application scenarios is preventing attacks when users are holding cloud telecommunication. In the remote interaction scenarios, how to guarantee the continuous-verifying authentication security and do not interrupt the running interaction? The biometric that is improved by RNN may be an ideal solution. Besides, some biometric authentication approaches are harder to forge and impersonate than traditional biometric methods, such as keystroke authentication and mouse movement authentication. These implementations should depend on the stable inherent characteristics of users that are collected by high-sensitivity biometric. RNN or LSTM RNN, these kinds of neural networks which are capable of remembering some prior history to greatly enhance the performance of biometric. Besides, RNN effectively improves the stability and reliability of biometric when they are under adversarial attack, because RNN is one of relatively mature models to deal with adversarial robustness tasks and anomaly detection.

The paper shared literature review to mention some novel applications of RNN in biometric. One of the most attractive contributions is recognizing users by filtering unique walking patterns. The authentication could be implemented by recording and recognition of inertial gait. RNNs learn the records which are saved by gyroscope and accelerator. RNN has better capability to filter key hidden characteristics from complicated high dimensional data which are hard to be extracted by traditional biometric and measurements. This paper provides me a lot of significant inspiration to explore better machine learning tools to improve biometric. The data collection often meets a lot of unpredictable and intractable problems. For example, how to efficiently sample key features from raw data is a big problem, because one of the key features of biometric is high distinguishing. In other words, the collected features must be unique and hard to be forged. Clearly, some extracted features by RNN are intractable to explain and hidden in some latent space. Good concealment of extracted characteristics and efficient and low-cost collection manners are promising and encouraging for development of biometric in future.

\cite{chen2019adaptive} illustrates a promising method to implement iris verification and recognition better than traditional iris segmentation algorithms. It shows an architecture based on CNNs combined with dense blocks for iris segmentation. The paper leverages the power of a dense-fully convolutional network (DFCN) that is processed by some popular optimizer methods, for example, batch normalization (BN) and dropout. Besides, the public ground-truth masks of the CASIA-Interval-v4 and IITD iris database have some inherent defects. They do not consider the labeled eyelash regions. The paper labels these regions that occlude the iris regions by using the Labelme software package. It takes great experimental results based on the UBIRIS.V2 iris database except for the mentioned databases.

\begin{figure}[H]
\centering
\includegraphics[width=0.4\textwidth]{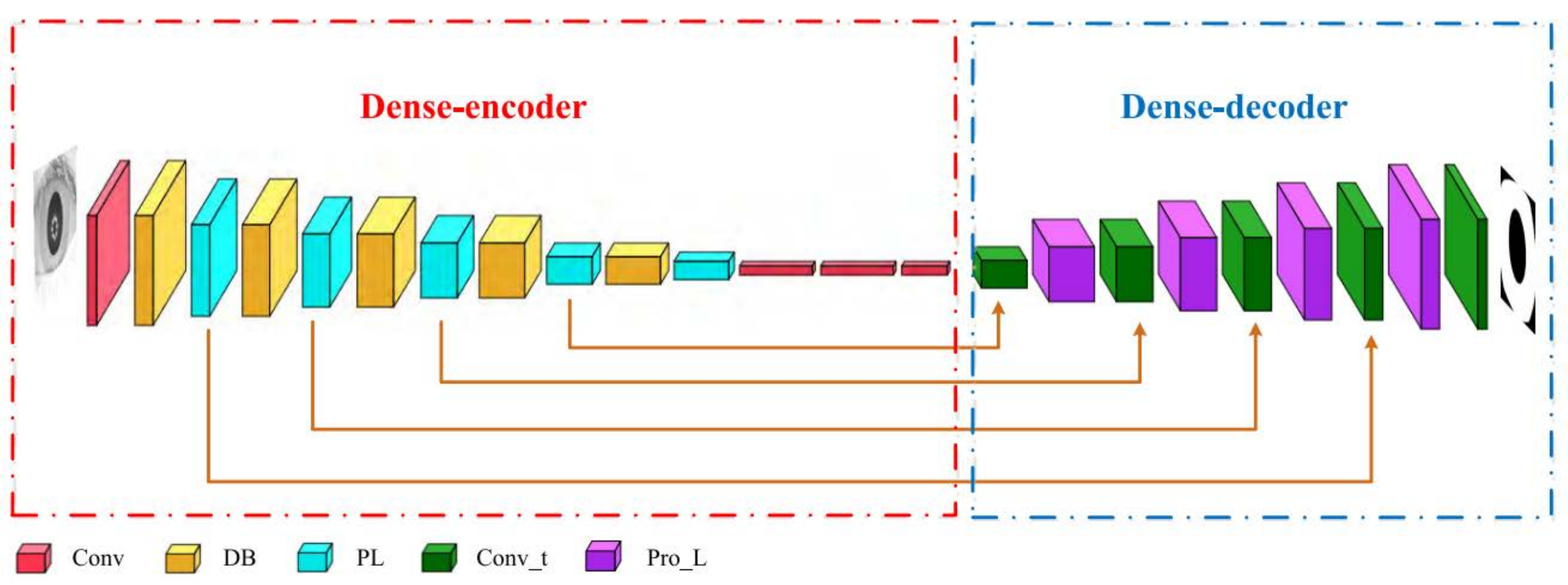}
\caption{The DFCN architecture.}
\end{figure}

The paper improves the iris segmentation by DFCN rather than conventional iris segmentation algorithms. Its proposed model shows outstanding performance on databases for different metrics. Besides, it proves that deep layers of CNNs should be a good option for enhancing iris segmentation.   
\begin{figure}[H]
\centering
\includegraphics[width=0.4\textwidth]{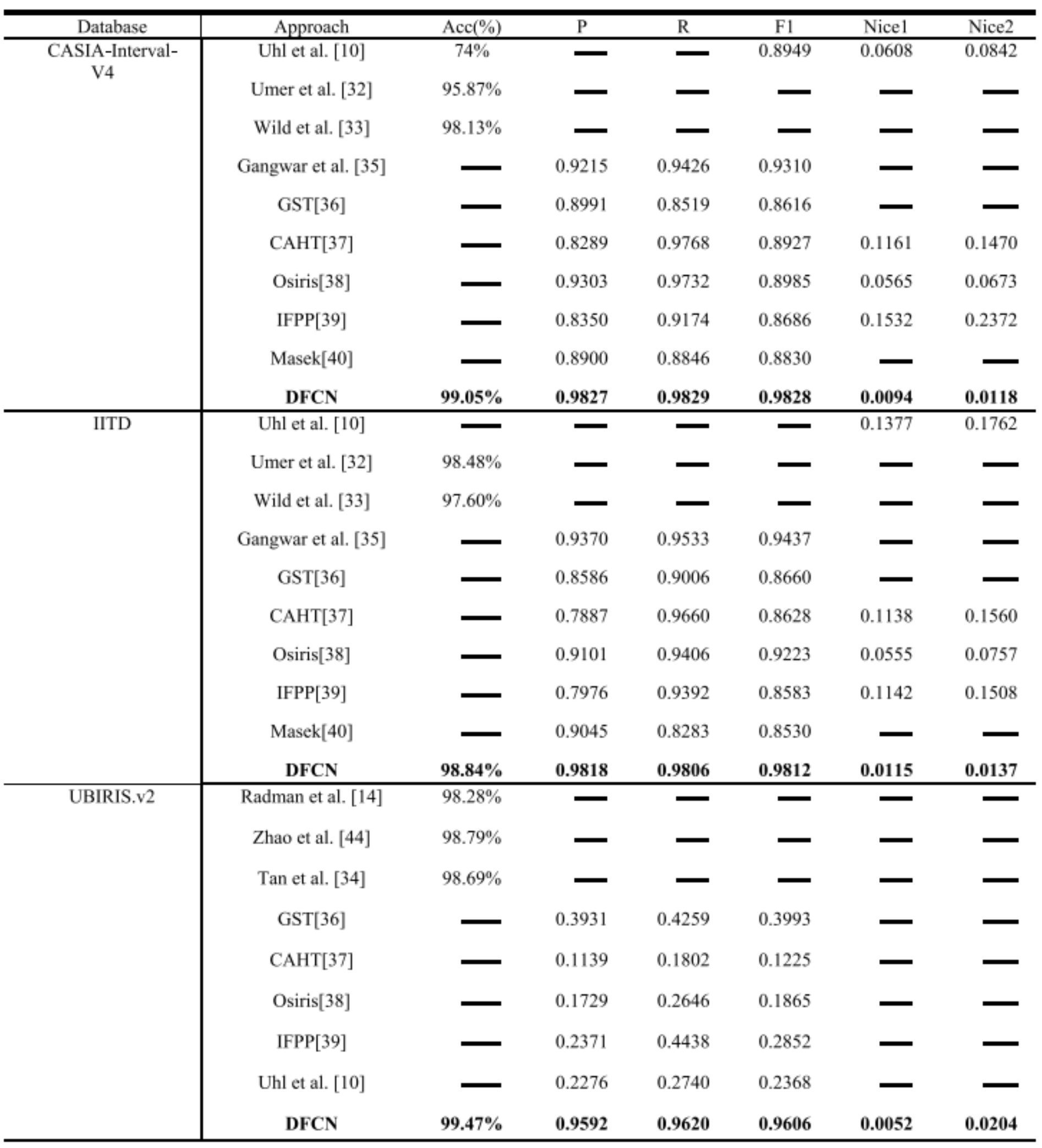}
\caption{Comparison of DFCN with conventional algorithms on the three iris databases.}
\end{figure}

\cite{ryu2021continuous} emphasizes securing systems have to face more challenges from various illegitimate access. Authentication plays a very important role to defend the security of systems. Continuous multimodal biometric authentication (CMBA) systems could secure the system effectively and that will be one of mainstream directions of development of authentication. This paper summarized current related studies and pointed out that most studies do not completely solve the problems on security, scalability and usability in practice.

Although most related studies show improvement on functional accuracy of CMBA systems, such as Equal Error Rate (EER) and False Rejection Rate (FRR), the models cannot reach ideal performance in a practical application environment. This paper found 23 percentage of papers tried to combine behavioral and physiological cues for an authentication system. It pointed out that each device could collaborate together to avoid the failure when one of the monitoring devices is out of order. But here we should try to avoid additional complicated and costly monitoring devices, because biometric requires universal for most customers. Schiavone adopted a special mouse that contains a fingerprint scanner to avoid interruption of user activity. Obviously, this innovation puts new requirements on development of portable and cost authentication biometric. This paper reminded us again, compared to traditional biometric, in future, authentication biometric should implement continuous and non-intrusive authentication schemes without additional devices, and do not depend on physiological characteristic-only measurement approaches.
\begin{figure}[H]
\centering
\includegraphics[width=0.4\textwidth]{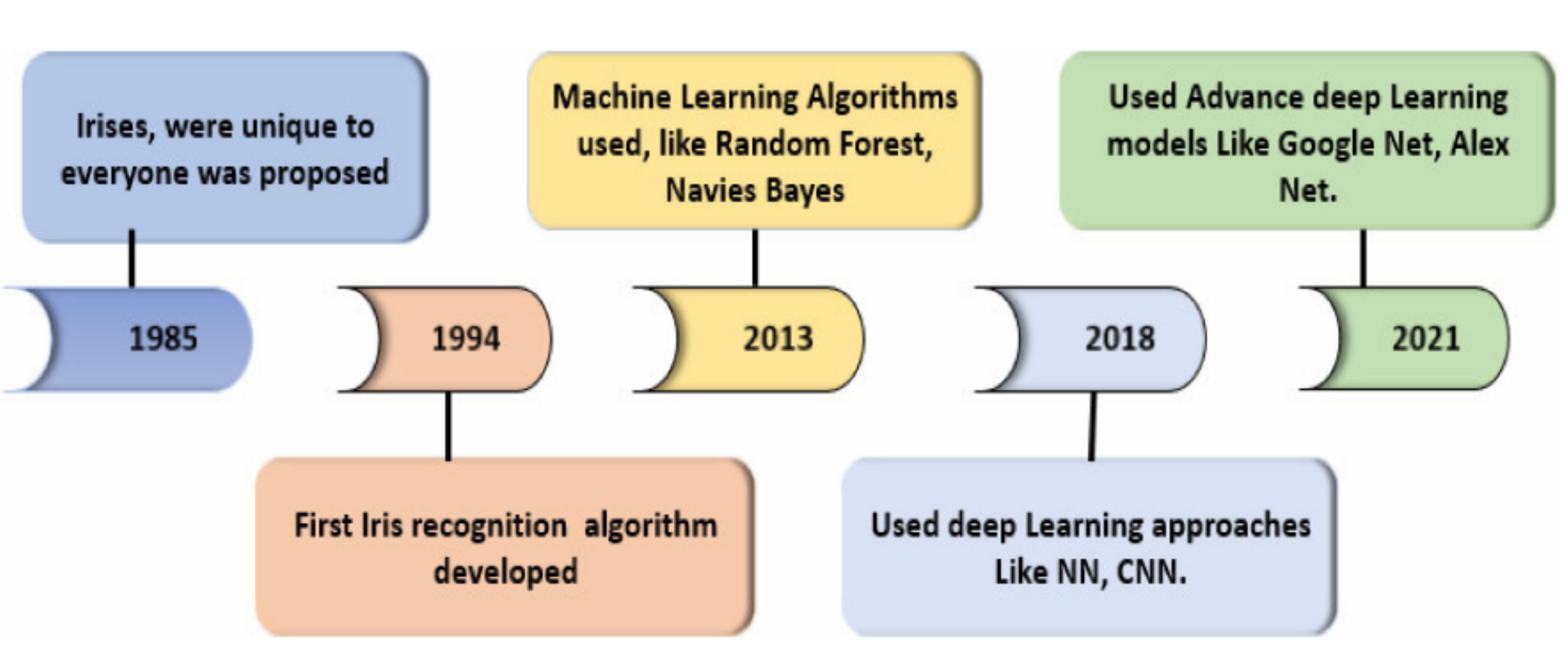}
\caption{Evolution of iris biometric identification.}
\end{figure}
\cite{khade2021iris} is a paper of literature review of iris liveness detection (ILD) for biometric authentication. It provides detailed and comprehensive investigation on related studies in this field. Nowadays, biometric is progressively vulnerable to professional attacks. Traditional Iris-based authentication biometric takes advantage of other biometric based on physical characteristics, because it effectively collects unique and permanent physical characteristics from iris through contactless verification. Besides, the production loss of Iris-based biometric devices of authentication has dropped sharply in recent years. However, some kinds of targeted spoofed attacks on Iris-based biometric have presented serious challenges, such as contact lenses, replayed video, and print attacks etc. Therefore, most related studies pay attention to improving the technologies of ILD.
\begin{figure}[H]
\centering
\includegraphics[width=0.4\textwidth]{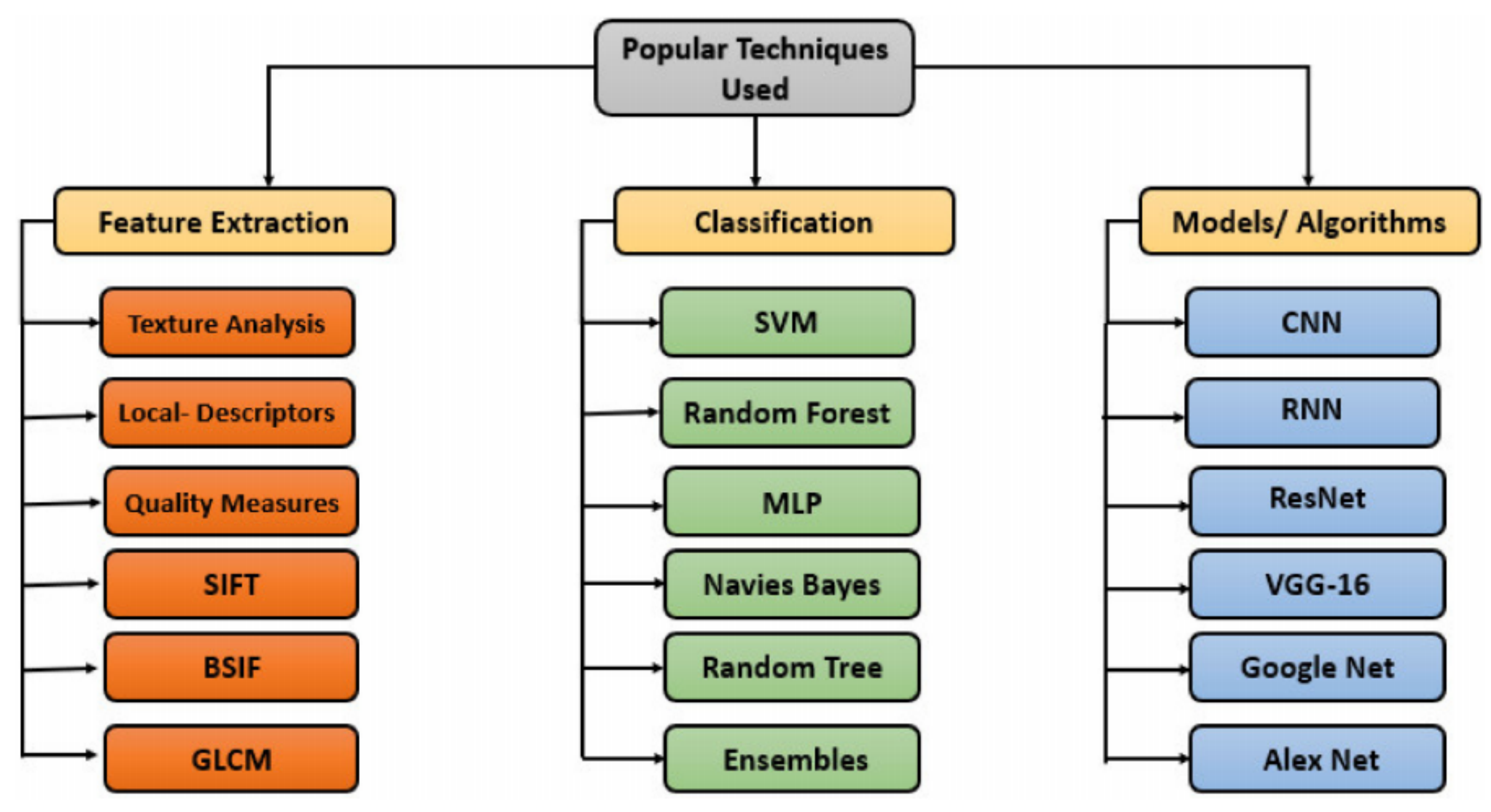}
\caption{Popular techniques used for iris liveness detection.}
\end{figure}

Systematic review of the paper pointed out machine learning and design-well data collection technologies greatly enhance the performance of ILD. The handcrafted features and self-learned features have good performance, but extracting handcrafted features are highly dependent on whether handcrafted feature extractors are proficient. Obviously, it is a big barrier to entry for the improved ILD based on handcrafted features. Clearly, self-learned features do not depend on proficiency extractors and professional extracting devices. So far, there are two major kinds of deep learning models for automatic extraction of Iris features. One is a regular deep learning model, the other one is a pre-trained model. In the field of machine vision, some pre-trained models show powerful capability for classification and identification tasks, such as VGGNet, ResNet and DenseNet, etc. Based on these models, transfer learning may be a good option to train them for Iris-based models. Figure 3 shows popular techniques used for Iris liveness detection. Deep learning models play dominant roles in models and classification tasks of Iris liveness detection. Deep learning will not replace the methods which are based on handcrafted features in a short time, but they are the main developing tendencies in this field.

\begin{figure}[H]
\centering
\includegraphics[width=0.4\textwidth]{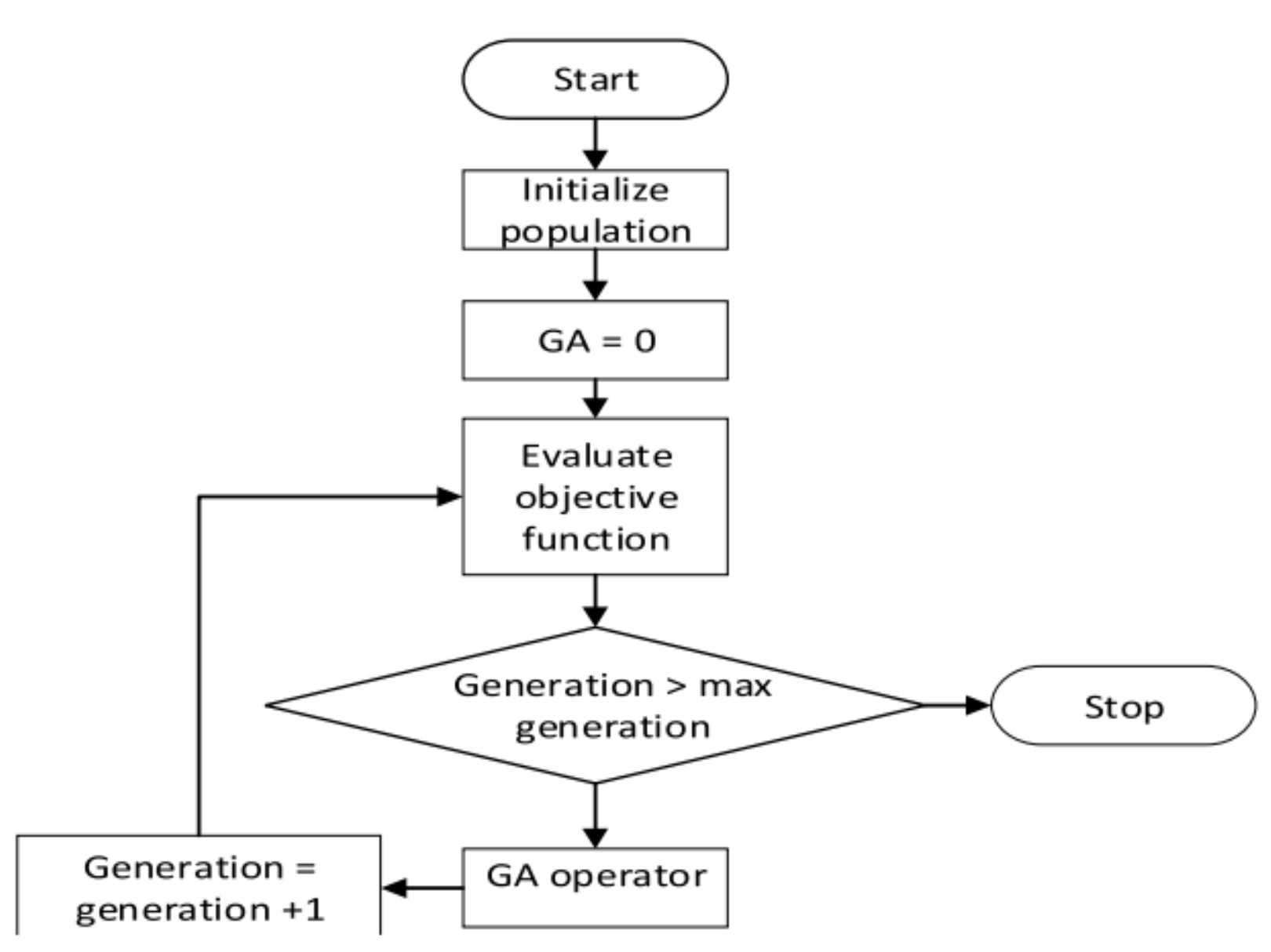}
\caption{Main steps of the Genetic algorithm.}
\end{figure}

\cite{el2021efficient} proposes a cancelable biometric framework which avoids attackers to forge the biometric template to reach the requirement of security. Its innovation is presenting a new authentication framework which adopts encryption techniques based on evolutionary Genetic Algorithm (GA). Thus, the framework generates a completely unrecognized and reconstructed biometric template. The significance of the biometric template which is processed by GA encryption technique is that attackers cannot explore the discriminative features from the template.

GA starts its research from a population of templates, but not only a single specific template. Besides, statistical operators play their roles to exploit the guiding the beginning population to produce successive populations. At last, the algorithm adopts crossover and mutation operations to generate ultimate cancelable biometric templates. The entire operational process is very hard to be reversed, if the operational seeds and schemes are not compromised. This paper provides significant inspiration for using encryption algorithms to improve biometric authentication. In most cases, one of the biggest challenges of the biometrics based on physical features is how to prevent attackers or hackers from forging and impersonating the templates of biometrics in a relatively short time. Once the template is duplicated, the entire biometric system will be in danger. However, the decryption algorithm is an optional solution to improve the performance of the biometrics in securing the template without any doubt. 

\cite{wu2021privacy} proposes a new method to integrate the fuzzy commitment and cancelable biometrics to promise a high level of security for biometric data. It has contributed to releasing a new scheme of privacy-preserving cancelable biometric authentication key agreement. The core of the scheme is dependent on a random distance method (RDM) which implements feasible cancelable biometrics and guarantees the templates of biometrics and data of users are not reconstructed. Besides, the pseudo-biometric identification and verification can be cancelable with privacy-preserving properties. After tests in experiments, the proposed scheme is considered to overcome most existing attacks through the mutual authentication of participants. Besides, the proposed approach has big advantages in computation and communication costs. This pros is really attractive for spreading the technique for cancelable biometrics. In practice, high accuracy of success rate of mutual authentication often means relatively high computation and communication costs. How to keep the high-level accuracy and high-efficiency authentication at the same time is a primary topic.

\begin{figure}[H]
\centering
\includegraphics[width=0.4\textwidth]{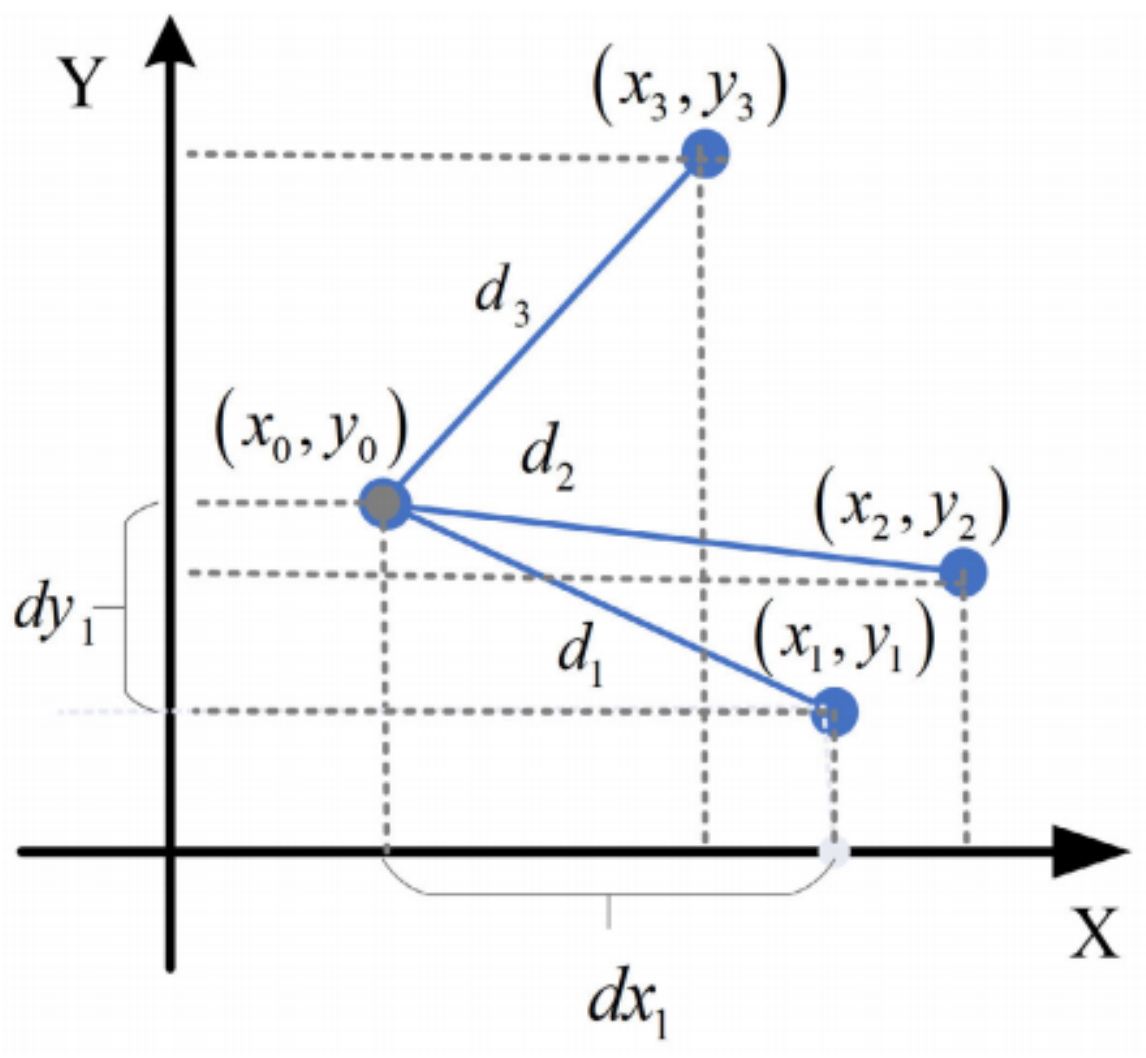}
\caption{The concept of random distance.}
\end{figure}
For template protection of the cancelable transforms of biometrics, Fuzzy commitment and secure sketch are often adopted and implemented by error-correcting codes. In this case, the distance between registration and probe biometrics needs to be determined by secured multi-party attempts by using computationally secure cryptographic tools. The advantages of RDM benefits the proposed scheme, because its computation cost is obviously lower than Random Projection (RP) which are often adopted in most current works. RP means the biometrics extract corresponding vectors of features and project them onto a specific random subspace. Its computation costs cannot be ignored in most cases. Thus, this paper inspires the innovations of biometrics authentication that could be based on improvement of algorithms to save computation costs and not sacrifice accuracy and security at the same time. This thought is undoubtedly significant and important.

\cite{ketab2022robust} introduces a novel application of biometric authentication for e-assessment. It requires the authentication mechanism to be robust, flexible, transparent and continuous. The paper adopted an eye tracker to detect cheating behavior through eyes’ communication. The participants of the exam need to pass through a process of registration to record critical facial information to generate a template of biometric. Then biometric automatically calibrates the fundamental movement of eyes when the participants stare at the screen or the monitor camera. This process only admits legitimate participants who pass through the identification. Then a 2D facial recognition mechanism will be activated and continuously identify the participants. During this time, their facial information will be uploaded and compared with the registration database. Every 4 second’s updating promises every inappropriate behavior will be recorded.

\begin{figure}[H]
\centering
\includegraphics[width=0.4\textwidth]{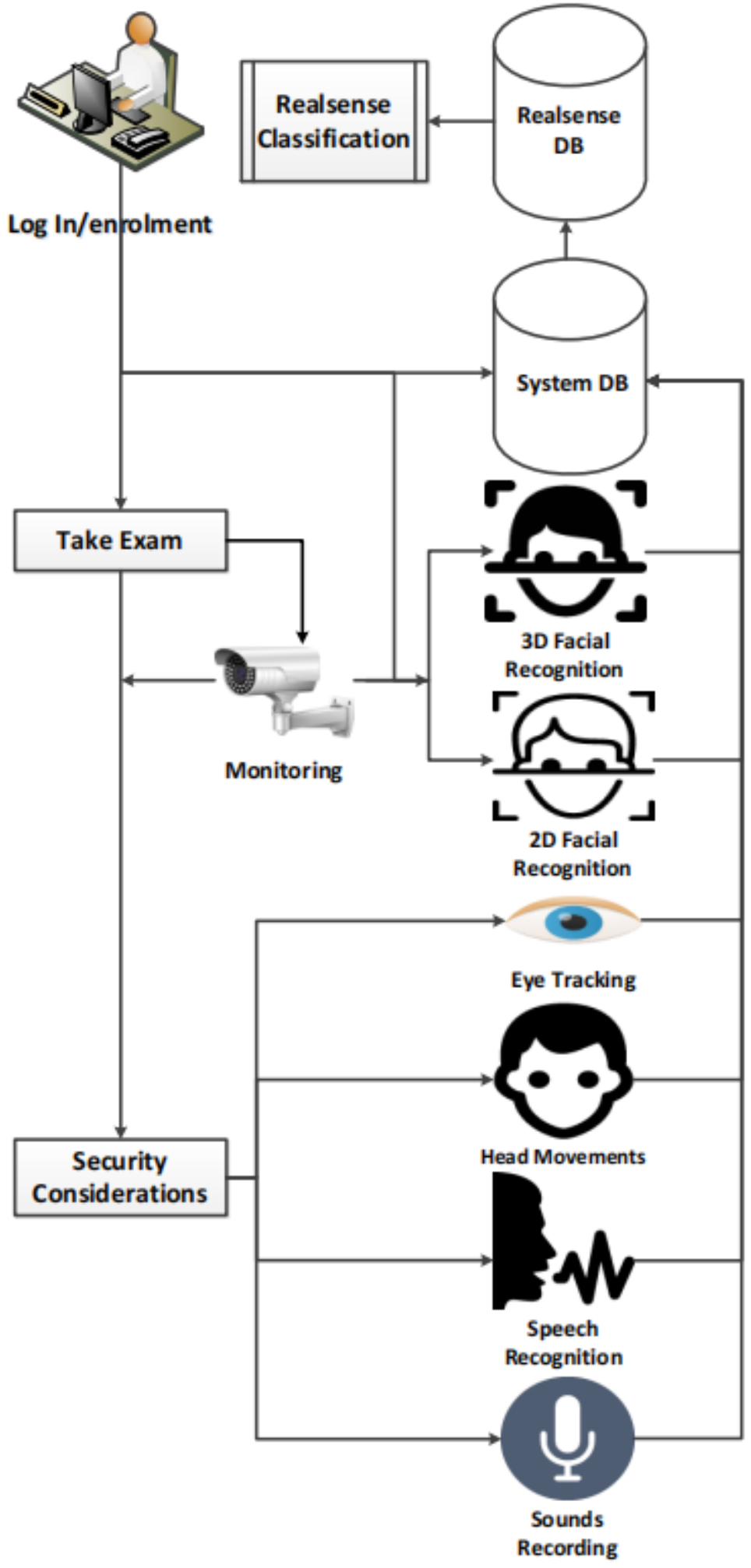}
\caption{System process diagram.}
\end{figure}

In practice, the camera only operates in 2D or 3D mode during specific time. After 5 minutes of the online matching assessment, the system switches the mode to continuous 3D facial identification. Successful matching operation and corresponding information will be uploaded to the system to calculate its 3D False Rejection Rate (FRR), failure matching operation will be alerted to measure the False Acceptance Rate (FAR). In this case, the eye tracker will be invoked and continuously working during either 2D or 3D modes. All the tracking facial information, such as movement of eyes and center locations will be recorded in this process. Besides, head movements are all recorded and measured to a text file. The paper provides a detailed scheme for biometrics to apply for e-assessments and biometric is inspired to spread in more application scenarios. 

\cite{osorio2021stable} proposes a privacy-preserving face identification system that adopts a Product Quantisation-based hash lookup table for indexing and retrieval of protected face templates. The hashes are useful to find the index of a face database. The templates are protected in the high privacy level of the enrolled subjects because they are defended by fully homomorphic encryption schemes. 

\begin{figure}[H]
\centering
\includegraphics[width=0.4\textwidth]{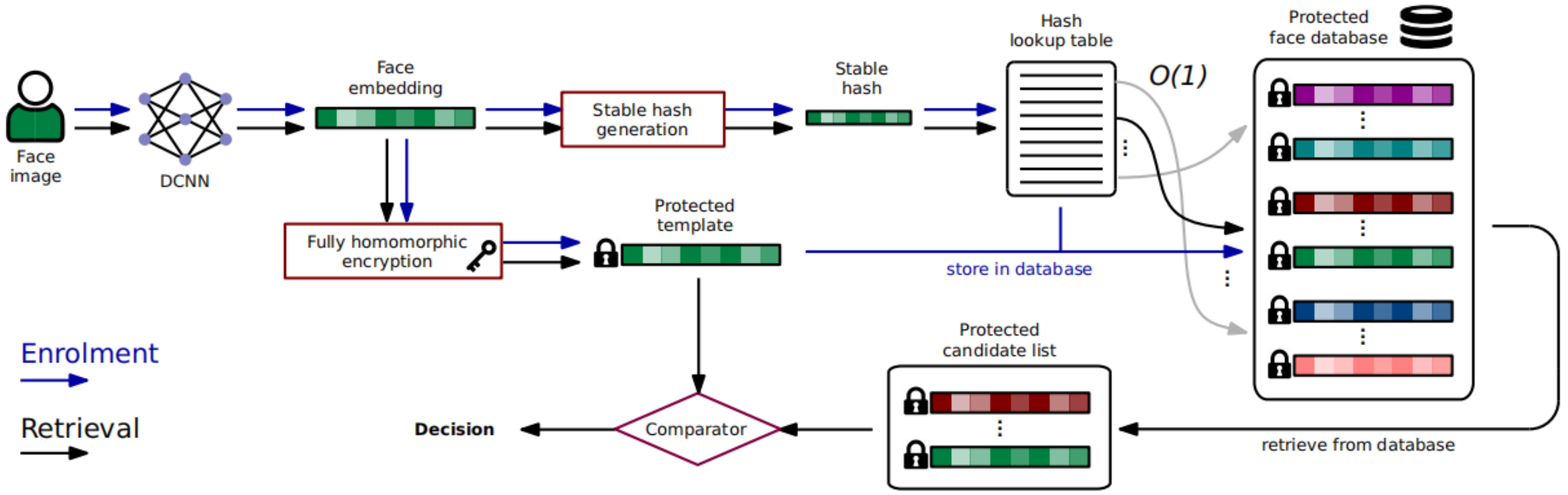}
\caption{Overview of a proposed privacy-preserving face identification system.}
\end{figure}

As the development of a large-scale facial identification system, the privacy protection of the enrolled subjects meets more and more severe challenges. Although there are several template protection algorithms that have been implemented previously. They are flawed more or less for indexing issues in biometric identification systems. In other words, the method adopts exhaustive searches for performing identification. The efficiency of implementation is relatively low. The paper greatly solves the problem. Its retrieval of the candidate short-list does not require a one- to many search. Instead of conventional methods, it can directly index through the hash lookup table. Thus, the time complexity of matching indexing is greatly improved to O(1). Its experimental evaluation with best configuration reduces the workload down to 0.001 of a baseline approach performing an exhaustive search together with a low pre-selection error rate of less than 0.01. 

\section{Conclusion}
According to a series of surveys of literature, biometric authentication has vast potential for future development. There are two main development approaches. The first direction is improved based on algorithms, such as improving machine learning, encryption algorithms to enhance the performance of biometrics. The other thought is improved based on data collection, for example, improving the portable and low-cost device to efficiently collect critical features for biometrics. Without any doubt, the future of biometric authentication is promising and limitless. It could imagine there will be more and more novel inspiring algorithms and collection approaches for enhancing  authentication biometrics.

\bibliographystyle{IEEEtran}
\bibliography{IEEEabrv,macros,UDCPS2022}

\end{document}